# From words to connections: Word use similarity as an honest signal conducive to employees' digital communication

Fronzetti Colladon, A., Saint-Charles, J., & Mongeau, P.

# From words to connections: word use similarity as an honest signal conducive to employees' digital communication

**Abstract**

Bringing together considerations from three research trends (honest signals of collaboration, socio-semantic networks and homophily theory), we hypothesize that word use similarity and having similar social network positions is linked with the level of employees' digital interaction. To verify our hypothesis, we analyze the communication of close to 1,600 employees, interacting on the intranet communication forum of a large company. We study their social dynamics and the "honest signals" that, in past research, proved to be conducive to employees' engagement and collaboration. We find that word use similarity is the main driver of interaction, much more than other language characteristics or similarity in network position. Our results suggest carefully choosing the language according to the target audience and have practical implications for both company managers and online community administrators. Understanding how to better use language could for example support the development of knowledge sharing practices or internal communication campaigns.

1. **Introduction**

The engagement of employees in an organization contributes to their wellbeing [1, 2] and to the organization's performance [3]. Employees engagement is realized through communication [4], which encompasses both the language people use to communicate and the interactions and relationships they have [5–7]. Nowadays, in most organizations, communication also happens through digital platforms [8, 9] (email, intranet, online forums), where employees can interact and exchange opinions and ideas [10]. Use of digital communication has been shown to predict employees' engagement level [11–13] and to impact future business performance [11, 14]. But what might foster employees' use of digital communication?

In recent years, several studies [11, 15–20] have used a network perspective to answer this question by studying patterns of communication as manifested in language use and interactions between employees. They seek to capture "honest signals" conducive to collaboration and communication among people [21], particularly signals appearing in digital conversations.

Adopting another stance, studies of socio-semantic networks have shown that interactions and relationships between people are linked to the similarity of the words and expressions they use, both face to face and online [22–25], while similarity in their attitudes, points of view and roles have been linked to their occupying similar positions in the organization's social network [26–29].

The present study proposes to consider similarity of word use and of social network position in order to contribute to a better measurement of "honest signals" conducive to digital communication between employees. Being able to capture these signals could enable the organization to foster effective digital internal communications and monitor their positive impact on the company performance.

**2. Honest Signals**

The concept of honest signal is borrowed from evolutionary biology [30]. A signal is an attribute or a behavior that changes the behavior of receivers to the benefit of the sender. An honest signal is one that reduces the uncertainty of a situation [31]: for example, the cry of a bird signaling the presence of a predator. Pentland [21] argues that the concept can be applied to human behavior to understand communication among employees within organizations. When we communicate with other people, there are subtle behavioral patterns that reveal our attitude towards them; these are unconscious social signals that complement our conscious language [21]. Indeed, our social behavior and use of language are the result of both conscious (intentional) and unconscious mental processes [32, 33], jointly creating "honest signals".

In their quest for honest signals favourable to collaboration and communication, Gloor and colleagues have considered three dimensions – language, social connectivity, and interactivity – that can act as signals for employees' participation in the digital conversation. In order to capture those signals, they used social and semantic metrics of digital communication and showed their signaling power in several business contexts [11, 15, 16].

Three aspects of language are included in the framework: complexity, sentiment and emotionality. Language complexity refers mostly to shared language: the more one uses dissimilar words (not used by others), the more one's language is considered complex. Sentiment expresses the positivity or negativity of the words used, whereas emotionality expresses the deviation from neutral sentiment or, in other words, the intensity of positive or negative feelings.

Social connectivity looks at an individual position within a communication network: the more central the individual is, the higher is his or her connectivity. Interactivity tries to assess how active individuals are in the network by analyzing the variation in their network position over time or by looking at how quickly (or slowly) they respond to posts, comments or emails.

In studies using this framework, authors have shown that low language complexity plays an important role in email communication with customers, ultimately improving their satisfaction [11]. In the same vein, the use of a shared language proved to support online community growth [20]. Hence, it appears that greater similarity in word use (less complexity) favors better and more active digital communication. Finally, language complexity and emotionality in the email communication of company managers were also shown to be linked to their level of engagement with their job [34].

Concerning interactivity, it has been shown that people's oscillations between central and peripheral positions in the network (called "rotating leadership" by the authors) plays a very important role in online communities, favoring their growth and participation [20], and that shorter response times to emails support better communication between clients and service providers [11]. There was also a significant association between rotating leadership and team creativity [35, 36]. Similarly, a higher level of rotating leadership proved to support the innovation capability of start-ups [37] and the probability that they would survive external shocks [38].

The results highlighted above show that the metrics for measuring the dimensions of language, social connectivity and interactivity uncover "signals" that can give managers invaluable information, leading to efficient action to support employees engagement. As maintained by Gloor [15] – who used these metrics to analyze digital communication in more than 200 organizations – considering together language use, social structure and interactivity, is necessary to obtain a complete view and comprehensively evaluate social dynamics in business contexts.

### 2.1. A Socio-semantic Perspective

With regard to language and interaction, studies have shown that interactions and relationships between people are linked to the similarity of the words they use [22–25, 39–42]. These studies use social network analysis to show how word use and relationships are intertwined.

More specifically, these studies are intended to explain and model the emergence of similarities between people's word use and the emergence, development and severing of social bonds. For example, Carley [43–45] argues that interactions can lead to a shared vocabulary and that word use is affected by changes over time in people's conceptual and social environments; Roth and Cointet [23] have studied how social and socio-semantic networks, two-mode networks linking people and the words they use, coevolve in networks of scientific collaborators and bloggers; Saint-Charles and Mongeau [25] have shown that centrality in an influence network is linked to word use similarity in workgroups and that this relation is transformed during the life of a group.

According to the theory of homophily [46, 47], similarity leads to the development of relationships based on attraction to others that are deemed similar to us [48]. Although most studies have explored similarities based on sociodemographic variables, several authors have extended their analysis to a wide range of variables, including attitudes, psychological traits, and values, that are seen as latent homophily factors [49–52], and have shown that the "homophily phenomenon" is complex and is not based solely on socio-demographic factors. Relational aspects, assortative mechanisms based on individual attributes and proximity factors can all influence the way people communicate and the frequency of their communication [37, 53–57]. As for network position, it appears that people occupying similar structural positions in the network [26–29, 58] tend to share similar opinions and behaviors. Maciel De Oliveira [59] shows that similarities between students occupying equivalent structural positions – and specifically a central position – enhance their tendency to identify with one another and to choose one another as workgroup partners. In a study of a large online network, Roy, Schmid, & Tredan [60] have shown that similarity in centrality is linked with role similarity of actors in the network. Finally, in an online learning community, Cho, Gay, Davidson and Ingraffea [61] showed that people move from peripheral participation towards full participation in the community of practice. In that process, the central actor's position was significantly correlated with the amount of information shared.

Given these findings, we assume that word use similarities between people, and particularly between people occupying similar centrality positions in a network, can act as an honest signal conducive to the development of social interactions in a digital communication network. Therefore, we hypothesize that:

H1. Employees' word use similarity is positively correlated with dyadic digital interactions.

H2. The similarity of employees' centrality in the digital communication network is correlated with dyadic interactions.

In other words, the more dyads are similar in their word use and their centrality position in the network, the more they will interact (or vice versa). Aside from their theoretical relevance in the search for an honest signal of collaboration and communication, these similarity measures should prove useful in the specific context of studying internal digital communication systems as they offer a way to identify clusters and inequalities in the distribution of interactions.

## 3. Research Design

In order to verify our hypotheses, we analyzed the digital communications of some 1,600 employees working for a large company. This company has an intranet social network, structured as an online forum in which only employees can interact, exchanging opinions and ideas as they share news and comments. In the style of well-known online platforms such as Reddit or the TripAdvisor travel forum, employees can either open new threads or comment on issues or news items that have already been posted. User posts generally discuss topics related to company performance or its internal and external initiatives – for example, employees comment on changes in HR policies, news of the company's performance in terms of earnings or stock prices, and news of the company's technological investments or its latest patent applications. The forum is also used to discuss work-related topics, with employees helping one another or sharing knowledge, for example in order to find solutions to technical problems. In this context, we were able to extract and analyze more than 23,000 posts (news and comments) written in Italian over a period of a year and a half. Corpus statistics are provided in Table 1.

| | |
|---|---|
| **Number of posts** | 23,031 |
| **Total number of words (Tokens)** | 2,440,467 |
| **Number of unique words (Types)** | 72,973 |
| **Type-Token Ratio** | 2,99% |
| **Words that occur only once (Hapax)** | 31,978 |
| **Hapax-Type Ratio** | 43,82% |

**Table 1**. Corpus statistics.

Users were mostly males (66% in our sample, also reflecting the percentage of male employees in the company) and a small proportion of them (7%) also acted as forum content managers. Content managers worked in the company's internal communication department and were responsible for shepherding online conversations, opening new topics, and answering users' questions. However, their assignment was not formal, so they were informally leading online discussion as content managers, without using an institutional writing style. Under agreed privacy arrangements, we are prohibited from revealing the company name or other details that could help in its identification. Data were processed in such a way as to protect employee anonymity: names were changed into numerical codes and message contents were not read (even though all messages were public on the company intranet). This is why we could not carry out a more in-depth content analysis, for example through topic modeling [62]. Reading messages was not necessary for the calculation of the semantic variables included in this study.

The first step in our analysis was to build a social network representing forum interactions. This network is made of N nodes, one for each forum user, and M edges. There is an edge between two nodes if the corresponding employees had at least one interaction – for example, they exchanged knowledge or opinions through comments, or one answered the other's question. We then proceeded to calculate similarity measures for both discourse and network centrality position. Figure 1 shows our network, excluding isolates and with node size scaled by betweenness centrality. Average degree is 18,78 and the average distance among reachable pairs is 2,26.

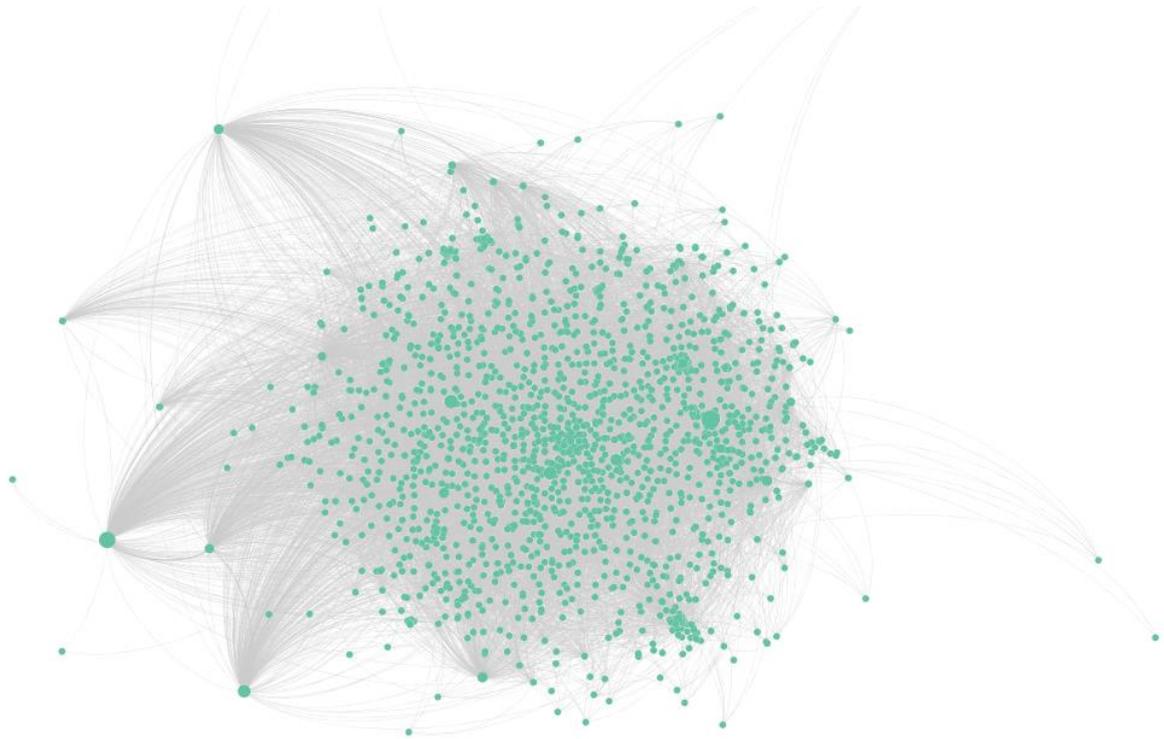

**Figure 1**. Interaction network.

### 3.1. Language Similarity Measures

Using the framework developed by Gloor and colleagues [11, 15] and the network-based studies on language and relationships cited above, we looked at five aspects of language similarity: word use (shared words between dyads), sentiment, emotionality, complexity, and length of each post.

*Text similarity* has been calculated using the Python programming language and the NLTK package. Original posts were preprocessed to put them in lowercase and to remove punctuation, special characters and stop-words – i.e., those words such as "the" or "and" that usually contribute little to the meaning of a sentence [63]. Subsequently, we also extracted stems by removing word affixes through the Snowball stemmer [64, 65] – this is an important step in dealing with the Italian language because many different affixes are used to distinguish between plural, singular, masculine and feminine forms, but the root word remains the same. Lemmatization would be an alternative approach, which can work well with languages like Italian that are morphologically richer than English, when an adequate vocabulary is available. After preprocessing, we transformed the posts written by each employee into tf-idf vectors (applying L2 normalization to take into account the different length of documents) and then calculated cosine similarity scores for each pair of users [66, 67]. Tf-idf vectors are commonly used in text analysis in order to attribute greater importance to words that can better characterize a text document. Specifically, words that frequently appear in a document, but are not frequently used in all other documents, obtain a higher weight [68]. Following this procedure, very common terms are almost filtered out in the process of identifying similarities in employees' word use.

We also investigated employees' use of language by looking at the *sentiment*, *emotionality*, *complexity* and *length* of their forum posts. Length is simply calculated as the average number of characters used in forum posts by an employee once stop-words and punctuation have been removed via a script written using the Python programming language and the NLTK package [63]. Sentiment (positivity or negativity of forum posts) is calculated with the machine-learning algorithm included in the Condor social network and semantic analysis software [15]. For the English language, this algorithm was trained on the basis of a large set of pre-classified text documents extracted from Twitter, as detailed in the work of Brönnimann [69]. For the Italian language, the developers of Condor followed a similar procedure and extracted a first training set from Twitter. Subsequently, we collaborated with them to refine the first version of their classifier and improve its accuracy in our research setting. For this purpose, we extracted 1,000 random intranet posts (of a length consistent with the average length of all posts) and asked two independent annotators to classify them as positive, neutral or negative. Initial ratings were consistent in more than 91% of cases. The two annotators subsequently met to review and agree on the discordant ratings. The classified documents were used to extend the original training set and improve the classifier's accuracy. Sentiment varies between 0 and 1, where 0 represents a totally negative post and 1 a totally positive one.

Emotionality expresses the deviation from neutral sentiment and is computed by Condor using the formula introduced by Brönnimann [70]:

$$\text{Emotionality} = 2 \cdot \sqrt{\sum_{i=1}^{n} \frac{(0.5 - Sentiment\ (W_i))^2}{n}}$$

where *Sentiment ($W_i$)* is the value of the sentiment calculated for any of the *n* words that appear in a single post. Posts containing words that are either strongly positive or negative are considered highly emotional. Emotionality has been frequently used in past research analyzing employees' communication; it has been shown to be significantly associated with employees' performance and engagement levels [16, 34].

Lastly, complexity represents the deviation from common language and is calculated based on the probability that each word will appear in the text based on the term frequency/inverse document frequency (TF-IDF) information retrieval metric [70]:

$$Complexity = \frac{1}{n} \sum_{w \in V} f(w)\ log \frac{N}{n_w}$$

where *n* is the total number of words in a post, *V* is the vocabulary of words that appear in all intranet posts, *f(w)* is the frequency of word *w* in the post, N is the total number of posts and $n_w$ is the number of posts that contain the word *w*. When rare terms appear more often in a forum post, its complexity is higher. This last measure was also calculated through Condor. Among the many software options, we chose Condor for semantic and network metrics as this software proved to be useful and reliable in past research [11, e.g., 15, 34]. Moreover, its development has been ongoing for many years.

We averaged scores of sentiment, emotionality and complexity to obtain metrics at the individual level (we considered all the posts written by each user).

**3.2. Network Position Similarity Measures**

To study the similarity of the connectivity and interactivity aspects of the framework proposed by Gloor [11], we used the indexes presented below.

To measure connectivity, we used network centrality, a construct commonly used in social network analysis to rank the position of social actors [71]. To measure centrality, we referred to the two well-known metrics of *degree* and *betweenness centrality*. Degree centrality measures the number of direct links of a node, i.e. the number of people an employee directly interacted with in the online forum [71, 72]. Betweenness centrality considers the indirect links of a node and counts how many times a social actor lies in between the paths that interconnect his/her peers. Betweenness centrality is calculated by considering the shortest network paths that interconnect every possible pair of nodes and counting how many times these paths include a specific employee (i.e. the node for which the betweenness centrality is calculated) [72].

Interactivity takes into account the evolution of the social dynamics over time. We operationalized interactivity by calculating *rotating leadership*. This variable counts the oscillations in betweenness centrality of a social actor back and forth on a weekly basis, i.e. the number of times his/her betweenness centrality changed significantly, with absolute variations of original values of at least 30%. Rotating leadership refers to informal communication leaders who move from central positions with high brokerage power to more peripheral roles (and vice versa) – thus permitting other employees to become central and mediate online interactions. The term "leader" is therefore meant to identify people who are central in the communication network, but it has nothing to do with formally appointed leaders [20]. The metric was calculated according to the procedure presented by Kidane and Gloor [35] and based on past research that considered weekly intervals appropriate for business contexts [11, 34]. We also tried changing the 30% variation threshold in betweenness centrality, without getting to better or significantly different results. On the one hand, if an employee maintains a static position, that person's rotating leadership is zero. On the other hand, we have rotating leaders when people oscillate between central and peripheral positions, activating or taking the lead in some conversations and then leaving space to other people in the network. The presence of rotating leaders has been shown to support the growth of online communities [20]. Rotating leadership dynamics also proved to favor collaborative innovation [73] and the performance of startups [37].

An additional metric often used to represent interactivity is the average response time taken by users to answer comments and questions directed to them. However, we could not compute this metric, as the timestamps associated with forum posts were not accurate enough. The timestamps were reliable with respect to the day/week, but not to the hour/minute, which is necessary to distinguish immediate answers from those which arrive several hours later. Measuring average response time in fractions of hours is indeed a common practice when analyzing digital communication in business settings; most answers usually appear within a few hours [11, 16, 34]. Accordingly, referring to day/week timestamps to calculate average response time would be highly inappropriate and would bias the results.

### 3.3. Control Variables

The control variables we could access were employees' gender and forum role (content manager or not). Even if gender homophily is not always supported by social network studies, it is very often used as a control variable, as it has been shown that gender can influence online social communication and behavior [74, 75]. Similarly, we control for content manager role, as we expect different behavior from employees responsible for informally moderating the intranet social network.

### 3.4. Similarity Matrices

Scores obtained for the above-mentioned variables were transformed into similarity matrices. Like a network adjacency matrix, a similarity matrix is made of N rows and columns, where each row and column represents a specific employee. For categorical attributes (gender and being a content manager or not) we have a value of 1 in a cell of the matrix if the two corresponding employees share the same attribute (for example they are both females), and 0 otherwise. For continuous variables, we populated the matrices with the absolute value of the differences in individual actor scores.

**4. Results**

To verify our hypothesis that the more dyads are similar in their word use and their network centrality position, the more they will interact (or vice versa), we first compared descriptive statistics and correlation coefficients for the social network and semantic variables.

We found that being a content manager was associated with more central and dynamic network positions: content managers had higher average scores of degree and betweenness centrality and they rotated more. In other words, they had interactions with more people, often acted as brokers of information, and in general did not keep a static dominant position after having fostered a conversation. Indeed, content managers often had the assignment of opening new forum topics for which they received comments that translated into incoming messages. It is also interesting to notice that posts written by central people had greater average lengths but used simpler language (with fewer rare words). These posts might have become more popular – and received many comments which pushed their authors to more central positions – because they were long enough to be informative, but still relatively easy to read and understand (less complex). Table 2 and 3 show the correlations and descriptive statistics of our social network and semantic variables at the actor level.

|   |                       | M                    | SD       |
|---|-----------------------|----------------------|----------|
| 1 | Gender                | 65.71% Male          |          |
| 2 | Role (Content manager)| 6.80% Content managers |        |
| 3 | Sentiment             | .657                 | .168     |
| 4 | Emotionality          | .327                 | .059     |
| 5 | Complexity            | 7.614                | .391     |
| 6 | Length                | 357.528              | 594.911  |
| 7 | Degree Centrality     | 17.765               | 64.018   |
| 8 | Betweenness Centrality| 4037.020             | 53311.098|
| 9 | Rotating Leadership   | 11.590               | 19.494   |

**Table 2**. Descriptive statistics.

|   |                        | 1       | 2       | 3       | 4       | 5       | 6       | 7       | 8       | 9 |
|---|------------------------|---------|---------|---------|---------|---------|---------|---------|---------|---|
| 1 | Gender                 | 1       |         |         |         |         |         |         |         |   |
| 2 | Role (Content manager) | -.175** | 1       |         |         |         |         |         |         |   |
| 3 | Sentiment              | .009    | .017    | 1       |         |         |         |         |         |   |
| 4 | Emotionality           | .013    | -.012   | .461**  | 1       |         |         |         |         |   |
| 5 | Complexity             | -.017   | -.079** | -.032   | -.091** | 1       |         |         |         |   |
| 6 | Length                 | .055*   | .056*   | .043    | .071**  | -.213** | 1       |         |         |   |
| 7 | Degree Centrality      | -.008   | .285**  | .045    | .113**  | -.280** | .390**  | 1       |         |   |
| 8 | Betweenness Centrality | .012    | .170**  | .050*   | .157**  | -.393** | .518**  | .887**  | 1       |   |
| 9 | Rotating Leadership    | .030    | .252**  | .031    | .060*   | -.173** | .214**  | .715**  | .538**  | 1 |

**Table 3**. Actor-level correlations.

Table 4 shows the Pearson's correlation coefficients of digital communications (network ties) with similarity metrics. To address the non-independence of network ties, we assessed the significance of coefficients through permutation tests based on the Quadratic Assignment

Procedure (QAP) [76]. As described in the previous section, we measured similarity with respect to several employee characteristics: their gender, role as content manager, word use, connectivity and interactivity. Dyadic text similarity shows the strongest association with digital communication (ρ = 0.48). Employees who more frequently used the same vocabulary communicated more with each other. Apart from gender and sentiment, homophily effects seem to be significant for all the other variables included in our study. Employees who were more similar with respect to their use of words, their interactivity and their degree centrality tended to interact more with each other. With regard to the role of content managers, we see a negative effect coherent with their assignment. In fact, content managers had an obligation to support online conversations, sharing knowledge about the company and answering employees' questions. Therefore, we might expect that they would interact more with other employees than with other content managers who had the same responsibility.

|    | Similarity Metric | 1 | 2 | 3 | 4 | 5 | 6 | 7 | 8 | 9 | 10 |
|----|-------------------|---|---|---|---|---|---|---|---|---|----|
| 1  | Network Interaction |  |  |  |  |  |  |  |  |  |  |
| 2  | Text | 0.475*** |  |  |  |  |  |  |  |  |  |
| 3  | Gender | 0.004 | 0.019*** |  |  |  |  |  |  |  |  |
| 4  | Role | -0.097*** | -0.173*** | 0.079*** |  |  |  |  |  |  |  |
| 5  | Sentiment | 0.004 | -0.003 | 0.013* | 0.013 |  |  |  |  |  |  |
| 6  | Emotionality | 0.051*** | 0.044*** | 0.015 | -0.001 | 0.204*** |  |  |  |  |  |
| 7  | Complexity | 0.140*** | 0.097*** | -0.001 | -0.060* | 0.017 | 0.101*** |  |  |  |  |
| 8  | Length | 0.142*** | 0.105*** | 0.024*** | -0.051* | 0.009 | 0.121*** | 0.348*** |  |  |  |
| 9  | Degree Centrality | 0.348*** | 0.332*** | -0.004 | -0.261*** | 0.012 | 0.148*** | 0.428*** | 0.434*** |  |  |
| 10 | Betweenness Centrality | 0.302*** | 0.242*** | 0.005 | -0.156*** | 0.026 | 0.197* | 0.546*** | 0.565*** | 0.892*** |  |
| 11 | Rotating Leadership | 0.267*** | 0.366*** | 0.012 | -0.226*** | 0.008 | 0.099*** | 0.246*** | 0.259*** | 0.717*** | 0.564*** |

**Table 4**. QAP correlation

We replicated the calculation of correlation coefficients of digital communication with similarity metrics, while filtering the network to compare content managers with other employees. This served to see whether the associations found in Table 4 were consistent within different groups. As Table 5 shows, correlations of digital communication with text similarity, degree centrality, betweenness centrality and rotating leadership remain fairly high and significant across both groups. These findings highlight the importance of such metrics, and suggest their potential role in digital communication. By contrast, homophily of role exhibits low coefficients, and similarity in gender and language sentiment remains negligible. Similarities in text length and emotionality are significant only for content managers, not for other employees. The effect of complexity is also sharply reduced for those who are not content managers. This suggests that, among employees acting as content managers, word similarity and the emotional content of language played a bigger role in shaping digital interactions. Content managers more often commented on the posts of colleagues who were aligned in terms of length, emotionality and complexity of the language they used. By contrast, these aspects were negligible when looking at the interactions of regular employees.

| Similarity Metric | Full Network | Content Managers | Non-Content Managers |
|---|---|---|---|
| Text | 0.475*** | 0.418*** | 0.436*** |
| Gender | 0.004 | -0.033* | 0.009*** |
| Role | -0.097*** | NA | NA |
| Sentiment | 0.004 | 0.060* | -0.007* |
| Emotionality | 0.051*** | 0.216*** | 0.003 |
| Complexity | 0.140*** | 0.266*** | 0.012* |
| Length | 0.142*** | 0.320*** | -0.003 |
| Degree | 0.348*** | 0.351*** | 0.212*** |
| Betweenness | 0.302*** | 0.337*** | 0.211*** |
| Rotating Leadership | 0.267*** | 0.297*** | 0.142*** |
| N | 1611 | 110 | 1501 |

**Table 5**. Network Intearaction QAP Correlations by Group

We completed the information provided by correlations with the multiple regression models presented in Table 6, in order to check whether digital communication could be at least partially explained by homophily effects. Accordingly, similarity matrices were used as input of QAP multiple regression models and we applied the Double Semi-Partialing permutation method to evaluate the significance of predictors [77]. Indeed, Dekker et al. [77] proved that this approach is one of the most robust against conditions of autocorrelation and collinearity. We also performed a preliminary calculation of the Variance Inflation Factor (VIF), which revealed significant collinearity only between the metrics of degree and betweenness centrality (max VIF = 8.41). In order to avoid collinearity problems in the QAP regression models, we performed a principal component factoring by combining these two variables and retaining just one single factor accounting for 94% of the variance, with the same factor loading of 0.97 for both degree and betweenness centrality. This new variable, named betweenness-degree factor, is higher when employees are more central in the online communication network. A similarity matrix was calculated consistently with the procedure described in the previous section, to express similarity of network positions between employees. QAP regression models were implemented using the R programming language and the Asnipe package [78].

Regression results confirm the preliminary findings of QAP correlations, highlighting possible homophily effects. We present our models following a hierarchical regression approach, i.e. adding blocks of variables to the initial model, to assess their impact on the adjusted $R^2$ coefficient. The first model includes word use similarity and text length similarity metrics (with word use being our main independent variable); the second adds characteristics of individuals (gender and the role of content manager); in the third, we add social network metrics; in the fourth we complete the analysis with semantic variables (sentiment, emotionality and complexity). Although all of our predictors are significant, except for sentiment, the effect size of many of them is negligible when compared with word use similarity and degree centrality similarity. This is also evident when comparing Models 1-4 with the more parsimonious Model 5. This last model has almost the same adjusted $R^2$ as the full model (0.2675 vs 0.2700), but only includes two predictors, text similarity and betweenness-degree factor.

| Similarity Metric | Model 1 | Model 2 | Model 3 | Model 4 | Model 5 |
|---|---|---|---|---|---|
| Text | 1.811660*** | 1.804603*** | 1.653310*** | 1.652887*** | 1.613879*** |
| Length | .000013*** | 0.000013*** | -0.000003*** | -0.000002*** | |
| Gender | | -0.001295*** | -0.001012*** | -0.001048*** | |
| Role | | -0.003498*** | 0.006507*** | 0.006775*** | |
| Betweenness-Degree Factor | | | 0.019318*** | 0.019907*** | 0.016154*** |
| Rotating Leadership | | | -0.000199*** | -0.000202*** | |
| Sentiment | | | | 0.001224 | |
| Emotionality | | | | -0.009983*** | |
| Complexity | | | | -0.003497*** | |
| Constant | -.034849*** | -0.030922*** | -0.032333*** | -0.031098*** | -0.029243*** |
| Adjusted $R^2$ | .2341 | .2342 | .2693 | .2700 | .2675 |

***p < .001.

**Table 6**. QAP multiple regression

To summarize, our results indicate strong homophily effects with respect to word use similarity. Employees tend to interact more with peers who share their vocabulary and a similar level of network centrality – i.e., central people interact more with central people and peripheral people with peripheral people. Our main findings are illustrated in Figure 2.

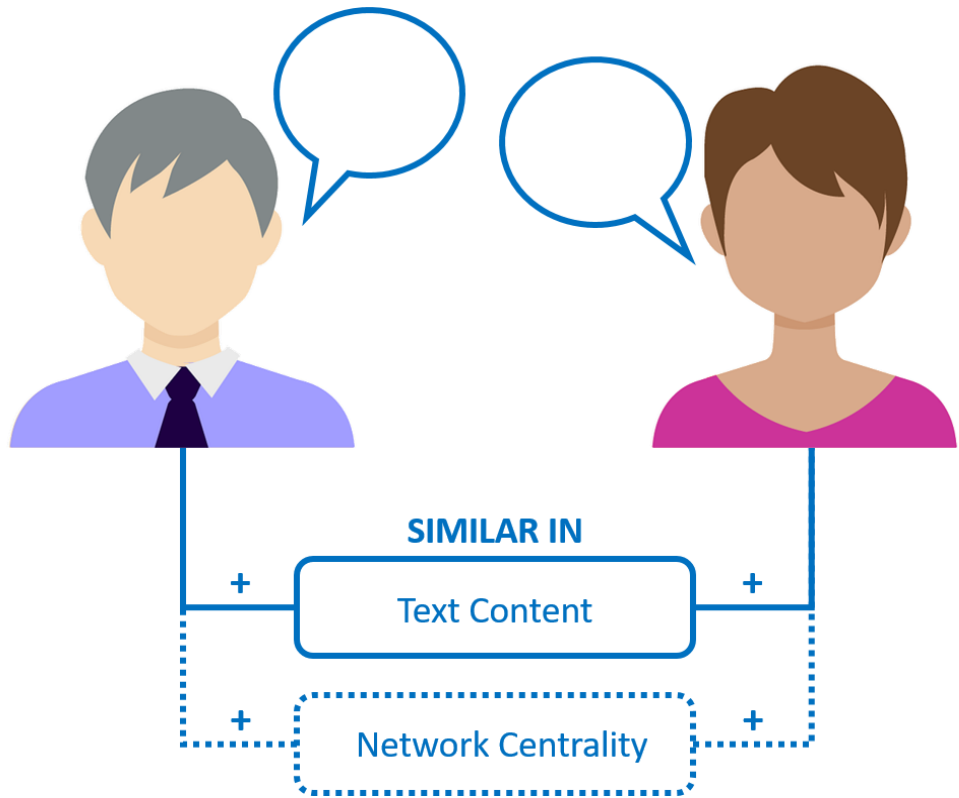

**Figure 2**. Main research findings (weaker effects are marked with a dotted line).

We also notice that more central people use a simpler language and write longer, perhaps more informative, posts. Lastly, content managers play an important role in building the online swarm [20], rotating more and being more central in social patterns.

5. Discussion

Overall, our results support and nuance our hypotheses stating that similarity in word use and in network centrality position between individuals is correlated with their interactions (H1 and H2). In our data, similarity of word use and centrality in the interaction network contribute to explain digital communication interactions taken as a dependent variable. Those employees who interact most with each other tend to share similar words; they also tend to have a similar number of links with others in the intranet forum and exhibit a similar level of betweenness-degree centrality.

However, those factors do not contribute equally to explain digital interaction. Indeed, although the full model explains 27% of the variance, a model with only text similarity explains 22.6% of the variance – a small difference leading to the conclusion that text similarity is the main driver of interaction.

Our findings differ from what Cho et al. [61] found in an online learning community in which the more an individual became central, the more he or she shared information with others (and therefore interacted with them). In our case, employees having similar centrality interact more with one another, which may lead to the creation of subgroups of people more or less central in the organization network, with various degrees of access to information and, potentially, various understandings of the organization' objectives. This may impede the development of organizational cohesion and a shared organizational vision.

Looking at results from the point of view of homophily theory, we may suppose that individuals are attracted to those whose word use resembles their own. To our knowledge, this is the first study combining social network and semantic analysis to show a strong link between the words used by employees and their digital interactions.

This research also extends the work of Gloor and colleagues, who framed the social and semantic metrics of digital communication and proved their signaling power in several business contexts [11, 15, 34]. These studies did not consider the homophily effects brought about at the dyadic level by employees' similarities. Actually, our results indicate that homophily, and especially language homophily, might moderate or mediate some of the results obtained in past research. For example, Gloor and colleagues [11] showed that faster responses to clients' emails, the presence of steady leaders, and a simpler language can all positively impact customer satisfaction. However, they did not explore the effects produced by language similarity between clients and the account team (which would be interesting to investigate in future research).

Our findings have practical implications for both company managers and administrators of online communities. For example, if a company wants to attract employees' attention to a strategic topic, in the light of our results, it appears vital to use words close to those used by the target group. Employees' participation in conversations can be fostered by online messages aligned with their use of words and by choosing social ambassadors who have degree centrality similar to the targets. From this perspective, the choice of the most appropriate ambassadors might be crucial, for example, for the success of internal communication campaigns carried out on intranet social networks.

## 6. Conclusion

Based on previous observations stating that the employees' use of digital communication can predict their engagement level [11], and that communication encompasses both the language people are using to communicate and their interactions and relationships [5–7], and in view of studies showing that interactions between people are linked to the similarity of the words they use [22–25], we hypothesized that similarity in language and in network position between individuals would be correlated with their interactions. Our results support and nuance our overall hypothesis showing that the main "homophilic" driver of employees' interactions is language similarity.

It might be useful to replicate our research to see if our findings are confirmed in different business contexts. Future studies could include more control variables, particularly those which are supposed to produce homophily effects, such as employees' age [79] or their geographical location (not available in our dataset). We know that all the employees involved in the study were working in Italy, even if they were grouped in different cities or buildings. The intranet social network was created by the company to support communication and knowledge sharing among geographically dispersed individuals. Indeed, past research has shown that geographical proximity can favor communication – especially face-to-face communication – but not necessarily the sharing of knowledge or of business-related information [37]. An analysis of the possible drivers of word use similarity is beyond the scope of the present study. For example, it could be that shared spaces impact language use and both digital and face-to-face communication [80]. This could be considered as control variable in future research.

As ours was mainly an association study, we advocate further research to carry out a longitudinal analysis, which could tell us which actor's similarity effects can be considered as significant antecedents of digital communication.

In addition, one could argue that having text similarity as one of the major drivers of digital communication is no surprise, as employees who comment on the same forum post will probably discuss a common topic, thus using similar words. This argument would apply to almost all of the studies about the overlap of social and semantic networks. However, in our study, this effect is at least partially mitigated by the fact that text similarity of employees is calculated with respect to their overall word use, i.e. considering all their posts, without looking at similarity within topics. Moreover, in an initial qualitative screening of the forum posts, we noticed that many employees maintained their own language, even when commenting on the same forum post. Therefore, word use similarity was only partially influenced by the fact of commenting on the same posts, and should be understood more as an overall metric of content similarity across different forum posts.

It is important to clarify that using the same words does not necessarily mean sharing the same opinion. Consider the extreme case where Person A says "The CEO is responsible for our drop in stock price and should be fired" and Person B says "The CFO is responsible for our drop in stock price and should be fired". Words use similarity is high, even if the two persons are saying different things. We do not see this as a limitation, as Person A and B are expressing different views in similar ways, i.e. using a common language. In this scenario, we classify A as much closer to B than to a Person C who says: "The CEO is totally incompetent: he is an idiot who should be kicked out of the company". In short, it is important to notice that we measured word use similarity and not agreement on business-related topics. Content analysis aimed at measuring agreement could certainly be interesting for future research (although we could not do it with the data used here, due to privacy agreements). This example highlights two important considerations: on the one hand, our study results support the link between language similarity and interactions shown by other studies in different contexts, none of which have considered agreement, indicating that similarity in the way opinions are expressed may be more important than the actual content of the opinions. For company managers and administrators of online communities, our results draw attention to the need to use similar words when one wishes to influence the opinion of others. On the other hand, it serves as a reminder that all interactions (or relationships), even strong ones, are not necessarily "positive" and calls for further exploration of the links between "positive/negative" interactions and language similarity.

Further research could explore the impact of the "word use homophily" effect on the overall network: does this lead to the creation of subgroups having different views of the organization? If so, how does it impact the organization's success? Our work could also be a starting point for future research that would see if the major drivers of digital communication on intranet social networks have the same influence in shaping relationships that take place on other media – email, phone calls, Skype calls, or face-to-face communication.